# Robust asynchronous optical sampling terahertz spectroscopy using commercially available free-running lasers

**Mayuri Nakagawa,[1,*] Natsuki Kanda,[1,2] Hidekazu Nakamae,[1] Hidefumi Akiyama,[1] and Ryusuke Matsunaga,[1]**

1. The Institute for Solid State Physics, The University of Tokyo, 5-1-5 Kashiwanoha, Kashiwa, Chiba 277-8581, Japan
2. Research Center for Advanced Photonics, RIKEN, 2-1 Hirosawa, Wako, Saitama, 351-0198, Japan

*Corresponding author: mayuri.nakagawa1@gmail.com

**Abstract:** This study presents asynchronous optical sampling (ASOPS) terahertz spectroscopy using commercially available Ti:Sapphire lasers without stabilizing repetition frequency. Our postprocessing algorithm using the multiplied repetition frequency difference as the calibration signal successfully corrected the jitter, thereby allowing broadband (2.5 THz) spectroscopy with a high spectral resolution (82 MHz). The robustness of the jitter correction based on the free-running laser setup and broadband electric circuits was rigorously examined under varying temperatures, thereby demonstrating reliable long-term operations over 60 h. This study expands the applicability of the ASOPS terahertz time-domain spectroscopy.

## 1. Introduction

Over the past few decades, terahertz (THz) waves and THz time-domain spectroscopy (THz-TDS) have been widely utilized to investigate the low-frequency electromagnetic responses in various fields, including solid-state physics [1–6], security [7], communication, pharmacy [8–11], and medicine [12–14]. Additionally, along with the advancements in optical frequency combs and dual-comb spectroscopy (DCS) [15–19], asynchronous optical sampling (ASOPS) using two femtosecond lasers has become a prominent method for data acquisition in THz-TDS, with high-resolution spectrum

[20,21]. The spectral resolution in THz-TDS is determined by the length of the temporal window, which is limited to the inverse of the laser repetition rate, $f_r$. In contrast to the conventional THz-TDS based on a single laser, the delay time between THz and optical sampling pulses in ASOPS is automatically scanned across the entire range of the THz pulse trains owing to the repetition frequency difference $\Delta f_r$ between two lasers, thereby eliminating the need for a mechanical delay line. The ultimate resolution as fine as $f_r$ in ASOPS THz-TDS can be utilized to detect narrow spectral features in gases and other molecules [22–25]. The long scan range is also advantageous for distance measurements [26]. The fast scan rate $\Delta f_r$ enables the observation of dynamics or moving objects within short frame rates [27]. Hence, ASOPS THz-TDS holds significant potential for future applications in security, pharmacy, medical, and physics investigations.

Despite these advantages, ASOPS THz-TDS has yet to be fully adopted in practical applications owing to limitations in the laser selection and measurement environment. In ASOPS and DCS, the choice of laser systems is crucial to suppressing a timing jitter, which arises from the instability of $\Delta f_r$. Jitter considerably hampers data accumulation in the time domain and reduces the measurement bandwidth, thereby necessitating effective suppression techniques. Conventional approaches include strict frequency stabilization control [22–24,28–33], specialized cavity design as single-cavity dual lasers [34–36], high-repetition-rate lasers [30], or incorporating CW lasers for adaptive sampling [36]. Contrarily, correcting the jitter during or after measurement offers a cost-effective solution to the jitter problem without requiring expensive experimental facilities. Jitter correction methods were initially developed for DCS [37–43]. In a report using two identical free-running pulsed lasers without additional laser [39], the DCS signal was calibrated by $10\Delta f_r$ signal extracted via radio-frequency filter, followed by correction based on the signal spectra. However, this method is challenging to apply directly to ASOPS THz-TDS because secondary self-correction by comparing single-scan signal spectra is not feasible for weak THz signals below the noise level. Further, the correction effect of the first calibration was insufficient for THz frequencies. Similarly, while arbitrary detuning methods in ASOPS employ interferograms for jitter correction [44], they are inadequate for THz-TDS.

Recently, we demonstrated a proof-of-principle of ASOPS THz-TDS using two independent pulsed lasers without any frequency control or extra lasers [45], which we refer to as jitter correcting (JC-) ASOPS THz-TDS. The residual timing jitter was reduced to 0.09 ps, an exceptionally small value than the other ASOPS jitter correction methods in other frequency regions with a large $\Delta f_r$. In the previous JC-ASOPS THz-TDS, homemade Yb-fiber lasers were used, while the jitter was corrected by recording a high-quality calibration signal $14\Delta f_r$. This method does not require precise laser control, thereby allowing for a greater flexibility in choosing the light source. Hence, a Ti:Sapphire laser is promising because it is one of the most widely used femtosecond pulsed lasers owing to its broad bandwidth and ultrashort pulse duration. Additionally, the emission frequency band corresponds with the characteristic direct-gap semiconductor GaAs, while commercially available chirped pulse amplifier systems can provide pulse intensities as high as several microjoules. This resulted in the development of efficient THz emitters and receivers based on Ti:Sapphire lasers. Consequently, implementing JC-ASOPS THz-TDS with a commercially available Ti:Sapphire laser system will improve the generation efficiency, detection sensitivity, stability, and measurement bandwidth of THz pulses, thereby broadening the applications of THz spectroscopy. Further, the jitter correction is potentially robust against rapid and considerable temperature variations. Hence, a closer examination of the tolerance of the JC-ASOPS THz-TDS to environment changes is required.

In this study, we developed a new system that provides sufficient calibration and demonstrated broadband, robust JC-ASOPS THz-TDS using commercially available Ti:Sapphire lasers. First, we described the principle of the JC-ASOPS system and presented the experimental results, including the system performance and calibration-free long-term measurement. High-speed, fine-resolution THz-TDS systems that are not limited to exclusive lasers or stable environments can expand the use and application of the THz measurements.

## 2. Measurement apparatus and signal processing

This section describes the optical and electrical systems used to acquire the ASOPS signal, triggering signal, and calibration signals. The software jitter correction method is explained below.

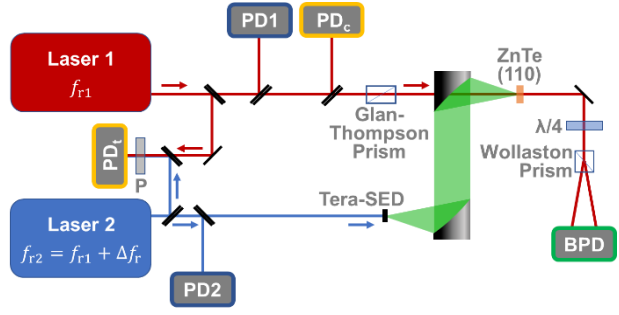

Fig. 1. Optical setup for the ASOPS using a pair of free-running pulsed lasers. P: polarizer; λ/4: quarter-wave plate; $PD_t$: photodetector used for interferogram detection to trigger the signal accumulation; PD1 and PD2: PDs for calibration signal generation; BPD: balanced photo detector used for ASOPS terahertz signal detection; $PD_c$: PD for digitizing clock detection.

The optical and electrical setups for the JC-ASOPS THz-TDS are shown in Figs. 1 and 2, respectively. Two free-running pulsed light sources were used for the THz pulse generation and electro-optic (EO) sampling with variable repetition frequencies $f_{r2} \sim f_{r1} \sim$ 82 MHz, and a repetition frequency difference $\Delta f_r \equiv f_{r2} - f_{r1}$ on the order of 100 Hz. These sources were commercially available Ti:Sapphire lasers (Tsunami, Spectra-Physics) placed on different optical tables meters apart. The time resolution of EO sampling is expressed as $|1/f_{r1} - 1/f_{r2}| = |\Delta f_r/(f_{r1}f_{r2})|$ and Nyquist frequency as $f_{\text{Nyq,EO}} = |(f_{r1}f_{r2})/2\Delta f_r|$. To maintain $f_{\text{Nyq,EO}}$ above 5 THz, $\Delta f_r$ must be kept below ~670 Hz. THz pulses were generated using a large-area interdigitated photoconductive emitter (Tera-SED, Laser Quantum), which provides high generation efficiency and is compatible with Ti:Sapphire lasers owing to the bandgap energy of GaAs. These THz pulses were detected in a (110)-oriented 1 mm-thick ZnTe crystal using a homemade balanced photodetector (BPD) optimized for fast ASOPS signals. The THz frequency signals are expressed as $f_{\text{THz}} = n_{\text{THz}}f_{r2}$, where $n_{\text{THz}}$ represents the natural number, which are down-converted to the ASOPS signal $(\Delta f_r/f_{r2})f_{\text{THz}} = n_{\text{THz}}\Delta f_r$ using an amplitude modulation by a gate pulse repetition rate $f_{r1}$. Because the repetition rate difference $\Delta f_r$ between the free-running lasers largely varies over time, the cut-off frequency of the BPD was designed to be as high as 100 MHz. In this setup, the bandwidth of the sampling digitizer (PXIe-5170R, National Instruments) was also 100 MHz. Because the cutoff frequency exceeded the gate repetition $f_{r1}$, the sampling rate of the digitizer was set to $f_{r1}$ to maintain the data point phase in the gate pulses. Practically, the sampling rate $f_{r1}$ was achieved using an external clock at a frequency of $2f_{r1}$ owing to the clock range of the PXIe-5170R. The external clock was generated using a photodiode ($PD_c$) that detected $2f_{r1}$ with a high signal-to-noise ratio (SNR), combined with a 110–180 MHz bandpass filter (BPF: ZABP-141-S+, Mini-Circuits) and an amplifier (ZFL-1000N+, Mini-Circuits), as shown in Fig. 2(a). An interferogram from two Ti:Sapphire lasers was detected to determine the time origin using a $PD_t$, DC–44 MHz lowpass filter (BLP-44+, Mini-Circuits), and DC-block (BLK-89-S+, Mini-Circuits), as shown in Fig. 2(b).

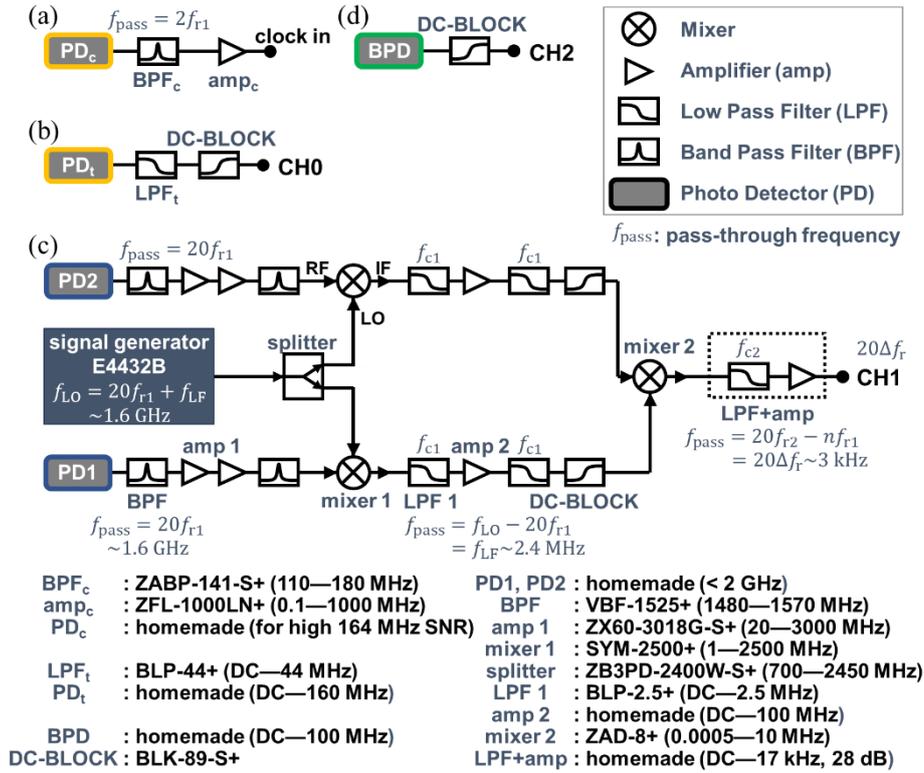

Fig. 2. Electrical setups for the broadband JC-ASOPS. Symbol descriptions are shown in the inset, with mini-circuits part numbers and bandwidths listed at the bottom. (a) Circuit for digitizer clock generation; (b) circuit for the accumulation trigger to record on channel (CH) 0; (c) circuit for calibration signal generation to record on CH1; (d) Circuit for the ASOPS signal to record on CH2.

The ASOPS signal was corrected using a calibration signal $20\Delta f_r$, generated by the electrical circuits shown in Fig. 2(c). Figure 3 shows the generation of the calibration signal $20\Delta f_r$ in the frequency domain. The calibration signal was generated from $20f_{r1}$ and $20f_{r2}$, detected by separate GHz range photodetectors (PD1 and PD2). The raw output from these detectors included a series of $nf_r$ values, where $n$ could range up to their bandwidths. Both PD outputs were roughly filtered at approximately $20f_r$ using 1480–1570 MHz bandpass filters (VBF-1525+, Mini-Circuits) with amplifiers (amplifier 1: ZX60-3018G-S+, Mini-Circuits). Further, the signals were shifted down to $f_{LF}, f_{LF} + 20\Delta f_r$ at approximately 2.4 MHz by taking a differential frequency with a $f_{LO} = 20f_{r1} + f_{LF}$ sinusoidal signal from a signal generator (E4432B/1E5, Agilent). The mixers (mixer 1: SYM-2500+, Mini-Circuits) performed a frequency addition/subtraction between $f_{LO}$ and multiple harmonic modes at approximately $20f_r$ to yield $\pm|f_{LO} \pm nf_r|$. However, only the lowest frequency component $f_{LO} - 20f_r$ passed through the lowpass filter with a cutoff frequency $f_{c1}$ =2.5 MHz (LPF 1: BLP-2.5+, Mini-Circuits). The second lowest frequency for each output after the mixer was $21f_r - f_{LO} = f_r - (f_{LO} - 20f_r) = f_{r1} - f_{LF}$ and $f_{r2} - (f_{LF} + 20\Delta f_r)$, which were higher than the cutoff $f_{c1}$. Hence, frequency components other than the lowest were removed. The lowest-frequency component was amplified (amplifier 2: ZX60-3018G-S+, Mini-Circuits) between two consecutive LPF 1, as shown in Fig. 2(c). Further, DC blocks (BLK-89-S+, Mini-Circuits) were added after LPF 1 to eliminate any slight offset from the amplifiers. The $f_{LO} - 20f_{r1}$ and $f_{LO} - 20f_{r2}$ signal after the DC-blocks were subtracted from each other to generate the calibration signal $20\Delta f_r$, using a low-frequency mixer (ZAD-8+, Mini-Circuits) and custom circuit that included amplifiers and lowpass filters (LPF+amp). The -3 dB frequency of the custom circuit was $f_{c2} \sim 17$ kHz. The Bode plot obtained prior to the JC-ASOPS is shown in Fig. 4.

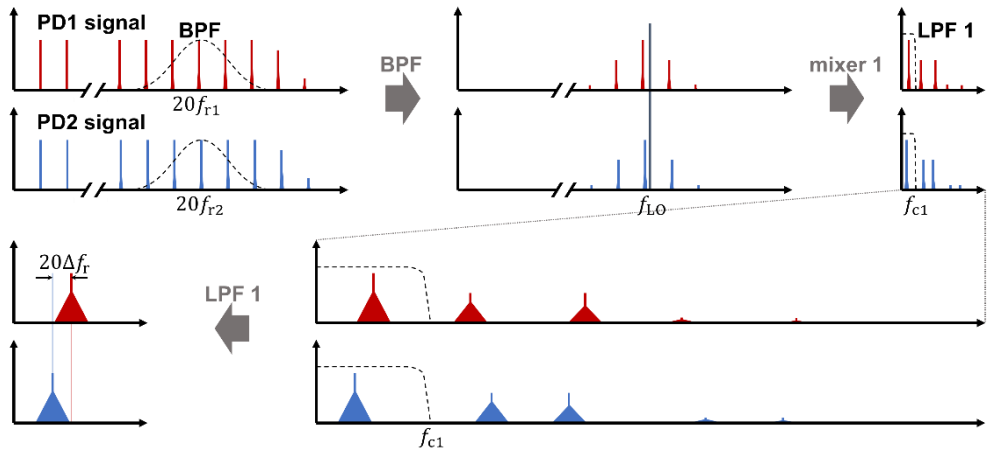

Fig. 3. Frequency-domain process of the calibration signal generation as shown in Fig. 2(c).

To prevent crosstalk from the adjacent signal $19\Delta f_r$ and $21\Delta f_r$, steep filters are required. However, to persistently generate a calibration signal for robust measurement in a changing environment, the output bandwidth must be sufficiently wide. A broadband calibration signal also crucial for correcting fast jitter. In methods that use the direct interference of comb-like spectra for calibration, as seen in self-correction techniques [38,39,42,46], the calibration bandwidth is inherently limited to $\Delta f_r/2$ owing to the spectral interval $\Delta f_r$ [47]. To mitigate this bandwidth limitation, the circuits as shown in Fig. 2(c) select the 20th signal and filter out other signals in the $nf_r$ signals before making interference between the lasers. Additionally, a signal generator was used to sharply filter adjacent signals by reducing the signal frequency. These configurations enable a steep filter with robust broadband measurements. The calibration signal can be generated as long as $\Delta f_r$ remains below $f_{c2}/20 \sim 850$ Hz, which is sufficiently high to account for potential room temperature fluctuations. For more unstable environments, the filter cutoff frequencies can be adjusted higher, thereby ensuring that the condition $f_r > f_{c1} > f_{LF} > f_{c2}/2 > 20\Delta f_r/2$ is maintained. Because $\Delta f_r$ is set to be below 670 Hz to keep $f_{\text{Nyq,EO}}$ higher than 5 THz for broadband THz spectroscopy, the requirement lower than 850 Hz for the broad-bandwidth circuit is naturally satisfied.

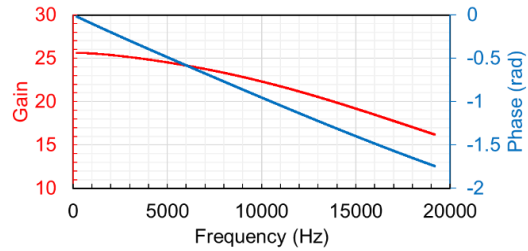

Fig. 4. Bode plot for LPF+amp as shown in Fig. 2(c). The red curve represents the gain in the linear scale. The blue curve represents the phase shift between the input and output.

Signal processing for jitter correction and accumulation was conducted using MATLAB or LabVIEW. Both software programs followed similar procedures, making the corrected time axis by interpolating the zero-crossing phases of the calibration signal after equalizing the falling and rising zero-crossing times by taking average. The intervals of the corrected time axis were eventually fixed by providing another interpolation of the ASOPS signal [45]. The phase plots as shown in Fig. 4 (blue) were used to compensate for the frequency-dependent delay in the calibration signal that occurred in the LPF+ Amp circuits as shown in Fig. 2(c). The compensation was achieved by shifting the corrected time axis, in the inverse direction of the frequency dependent delay [45]. The signal rearranged on the corrected times was accumulated and averaged to improve SNR. The time origin for the accumulation was determined by using the peaks of the interferogram as accumulation triggers. MATLAB was used for the lossless accumulation of continuous data after data acquisition, while LabVIEW was used for the successive accumulation of long-term measurements. In long-term measurement using LabVIEW, the accumulation was conducted within data acquisition, in lower rate limited by calculation speed.

## 3. Results

### 3.1 Continuous Measurement

Continuous measurements were conducted for 244 s using MATLAB to evaluate the accumulation performance of the system. For this measurement, the interchannel delay between the ASOPS signal and calibration signal was also compensated, in addition to the correction procedure described in Section 2. The interchannel delay compensation was determined by testing 100-time accumulations. The results are shown in Fig. 5(a). Accordingly, the optimal inter-channel delay which maximizes the peak amplitude was determined to -800 data points (-9.75 μs in laboratory time) and applied to the jitter correction throughout the entire measurement. All the 33,877 scans of the ASOPS signals with a data point length $\Delta f_r/f_{r1}$ were calibrated onto 1,000,000 points and accumulated into a single wave with the delay time range of $1/f_{r2}$. As such, small variations in scanned delay length $1/f_{r2}$ were disregarded, contrary to the correction for the variation in the scan rate $\Delta f_r$. Figure 5(b) shows the FFT power spectrum of the time-domain waveform after 244 s (33,877 times) of accumulation. Here, the data are normalized at their respective noise levels, which are determined by the mean power in the 5–7 THz, to show the SNR. The signal exceeded the noise floor by 2.5 THz, except for the absorption spikes caused by water vapor. Generally, to enhance the SNR in ASOPS measurements at high frequencies up to the THz region, a sophisticated laser is required, such as frequency-locked lasers, single-cavity dual-lasers, or CW THz lasers, in addition to the two-pulsed lasers. This is because the high-frequency components of the signal with a period shorter than the jitter vanish owing to accumulation, particularly when the time range for FFT is long (for fine frequency resolution measurement). In the previous JC-ASOPS studies, the first ASOPS THz-TDS was achieved using two independent free-running Yb-based fiber lasers without an extra laser cavity, although the observed THz signal bandwidth was limited to 1 THz for a frequency resolution of 100.78 MHz, with an accumulation time of 16 min [45]. The remarkably broadband ASOPS measurement of 2.5 THz as shown in Fig. 5(b) was achieved via efficient THz generation and detection using Ti:Sapphire lasers, combined with the high-performance jitter suppression by JC-ASOPS. As shown in Fig. 5(b), the frequency resolution is $f_{r2} = 82,035,907$ Hz and accumulation time is 244 s. For comparison, the average power SNR of the Ti:Sapphire measurement in the 0.95–1.05 THz range was 4405. The achieved SNR was 38 times higher than that of the previous Yb-based measurement, with an accumulation time 3.9-times shorter. This indicates that the result at approximately 1 THz was improved by approximately 150 times for a similar accumulation time. During the measurement, $\Delta f_r$ fluctuated (Fig. 5(c)) around its harmonic mean of 139 Hz. The varying sampling frequency $2f_{r1}$ was measured using a frequency counter, thereby allowing for the determination of the scan rate $\Delta f_r$ and resolution $f_{r2}$ based on the number of sampling points within each cycle of calibration signal $20\Delta f_r$. The harmonic mean of $f_{r2}$ was used as the frequency interval for the accumulated signal spectrum. The residual jitter on the corrected time axis was estimated from fluctuations in the trigger-to-trigger intervals to evaluate the correction performance. The root-mean-square of the residual jitter was 0.14 ps, while the jitter power spectrum obtained using the nonuniform FFT is shown in Fig. 5(d). The spectrum exhibits an $f^{-2}$ dependence on the jitter frequency, which corresponds with the white frequency noise. This measurement demonstrated a residual jitter smaller than half a cycle of 3.5 THz, even when using the free-running lasers with varying $\Delta f_r$.

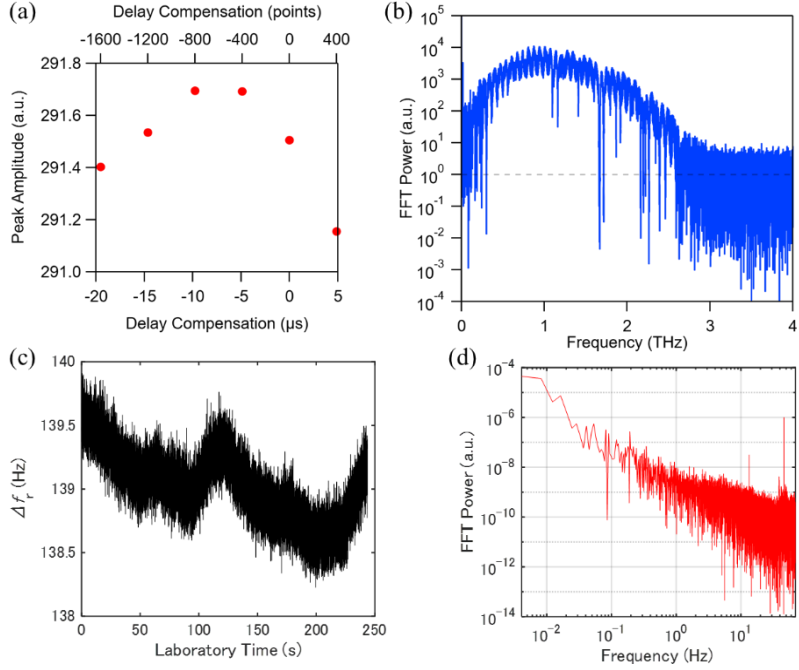

Fig. 5. (a) Peak amplitudes from 100 accumulation tests for various inter-channel delay compensations; (b) frequency spectrum of the waveform averaged over 224 s (33,877 scans) using JC-ASOPS; (c) $\Delta f_r$ during the measurement estimated from the intervals of the calibration signal; (d) power spectrum of the residual jitter in JC-ASOPS analyzed using nonuniform FFT.

*3.2 Long-term measurement*

One of the advantages of the JC-ASOPS is its ability to eliminate the limitations imposed by the control device in the conventional ASOPS, such as the operation range and bandwidth of a piezo actuator. To maximize the flexibility of jitter correction, the electric circuits in this study were selected to avoid restrictions on the $\Delta f_r$ frequency range for the calibration signal generation and enable broadband jitter calibration. In this section, the robustness of the JC-ASOPS against perturbations in the repetition frequency is assessed using a long-term measurement lasting 63 h. The repetition frequencies of two lasers were $f_{r1} \sim f_{r2} \sim$ 82,024,000 Hz, with an initial offset $\Delta f_r \sim$ 166 Hz. This configuration ensured that the frequency of calibration signal remained sufficiently high to correct for a fast jitter, even if $\Delta f_r$ varied during the measurement. The waveforms saved and processed using LabVIEW consist of averaged signal obtained after accumulating 1000 corrected waveforms. The accumulation was applied to the calibrated data, which were reduced by 1/36 owing to the calibration speed. Figure 6(a) shows a two-dimensional plot of the THz pulse waveforms for 1304 successive measurements.

The horizontal axis corresponds with the corrected time relative to the trigger, ranging -10–30 ps from the peak of THz pulse in the first accumulated scan. Figure 6(b) shows the spectrogram obtained using the FFT for the corresponding time window. The first and last scans of the accumulated time-domain waveforms and their FFT spectra are shown in Figs. 6(c) and 6(d), as represented by red and black plots, respectively. This excellent agreement between these plots demonstrates the superior robustness of the JC-ASOPS THz-TDS over 60 h.

Figure 7(a) displays the temperature variation near laser 1 (red) and laser 2 (blue) recorded concurrently with the ASOPS measurement in Fig. 6. The maximum temperature changes were 0.9 and 1.2 °C, respectively, with differing timing and magnitude, and the difference was potentially larger inside the laser enclosures. Figure 7(b) shows the trace of $\Delta f_r$ estimated from an expansion ratio between the laboratory time and corrected time during the measurement. A considerable drift in $\Delta f_r$, -157.7–-221.7 Hz, was observed owing to temperature variations, with a notable rapid drift occurring after 58 h. Despite this considerable frequency drift in $\Delta f_r$, the measured THz pulse data shown in Fig. 6 remained quite stable for 63 h in the time-waveform correction and spectrogram. Notably, the peak times of all the accumulated waveforms shown in Fig. 7(c) were at the same point, thereby indicating that the jitter in the accumulated data remained consistently below the time resolution of approximately 0.012 ps. Figure 7(d) shows the mean FFT power of the 0.95–1.05 THz signal, which remained stable even when viewed on a linear scale. The robustness of the jitter correction demonstrated in this experiment enhances the sensitivity through long-term accumulation and holds significant potential for applications such as continuous absorption monitoring.

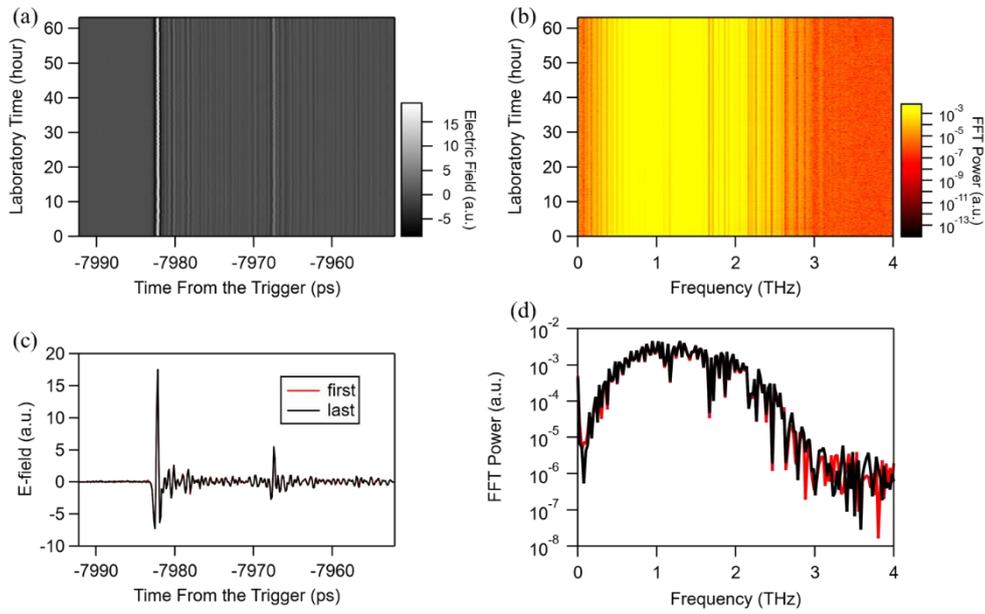

Fig. 6. (a) 2D plot of the 1304 THz waveforms measured over 63 h, wherein each waveform was averaged 1000 times after jitter correction; (b) spectrogram corresponding with the waveforms in (a); (c) time-waveforms extracted from (a), with the first (red) and last (1304th, black) waveforms shown for comparison; (d) frequency spectra extracted from (b), thereby showing the first (red) and last (black) measurements.

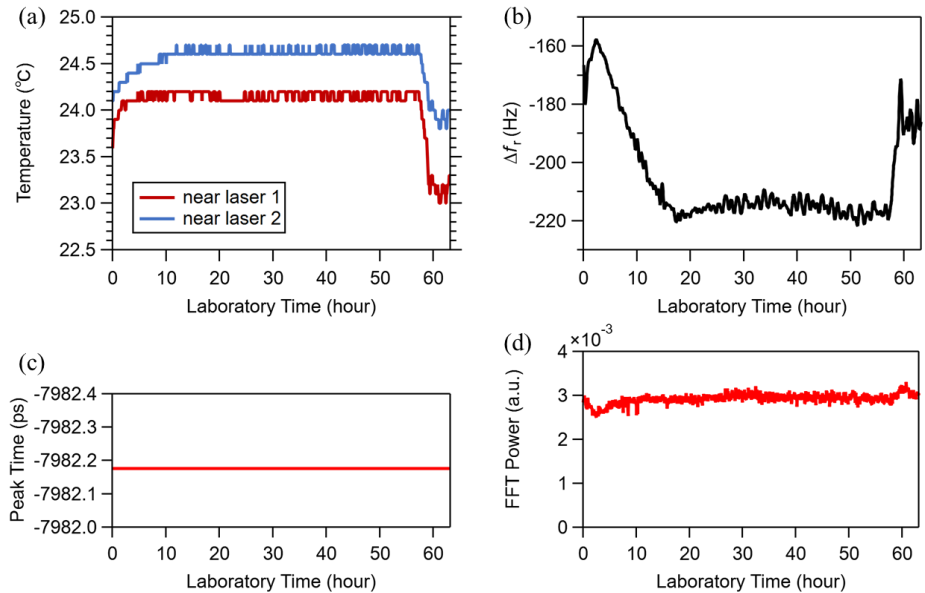

Fig. 7. (a) Temperature variations during the long-term measurement. The red and blue curves indicate the locations of the thermometers as specified in the legend; (b) $\Delta f_r$ trace during the long-term measurement estimated from calibration signal intervals; (c) peak times relative to the time origin in the measured terahertz waveforms; (d) mean spectrum power at approximately 1 THz (0.95–1.05 THz) during the measurement.

## 4. Conclusion

In this study, JC-ASOPS measurements were conducted using commercially available Ti:Sapphire lasers. Electrical circuits optimized for the broadband jitter correction and flexible generation of the calibration signal realized an ASOPS THz-TDS up to 2.5 THz with a fine frequency resolution of approximately 82 MHz. Further, the results also demonstrated robust calibration over 63 h using free-running, calibration-free laser operation, even with a considerable $\Delta f_r$ fluctuation of 64 Hz during the measurement. The residual jitter after the correction was 0.14 ps and less than 0.012 ps after 1000 accumulation. This exceptional stability was achieved through the calibration signal with a frequency of, $20\Delta f_r$, while feasibly low, remained sufficiently high to effectively monitor and correct the jitter. This study is the first THz-range free-running ASOPS using Ti:Sapphire lasers, while this method will broaden the potential application of ASOPS and THz-TDS by mitigating the limitations related laser selection, environmental stability, and accumulation time. Notably, the sensing of moving objects can be made more feasible by utilizing the fast scan rate of the ASOPS and an optimal laser to improve sensitivity. Additionally, this method enables a wider range of pump-probe terahertz spectroscopy applications by utilizing improved scan rate, high-frequency resolution, stable operation, and flexible and individual laser selection.

**Funding.** Japan Society for the Promotion of Science (JSPS) KAKENHI (JP22K18269, JP24K00919).

**Disclosures.** The data for this study are available from the corresponding authors upon request.

**Data availability.** The data for this study are available from the corresponding authors upon request.

**Conflict of interest.** The authors declare no conflicts of interest.